\documentclass[11pt,a4paper]{article}
\usepackage{amsmath}
\usepackage{amsfonts}
\usepackage{epsfig}
\usepackage{times}
\newcommand{\be}{\begin{equation}}
\newcommand{\ee}{\end{equation}}
\newcommand{\ba}{\begin{array}}
\newcommand{\ea}{\end{array}}
\newcommand{\bea}{\begin{eqnarray}}
\newcommand{\eea}{\end{eqnarray}}

\newcommand{\rar}{\rightarrow}
\newcommand{\p}{\partial}
\newcommand{\ol}{\overline}
\newcommand{\ti}{\tilde}
\newcommand{\la}{\langle}
\newcommand{\ra}{\rangle}

\renewcommand{\l}{\newline\null}

\def\hbar{h\!\!\!/}
\begin{document}
%%%%%%%%%%%%%%%%%%%%%%%%%%%%%%%%%%%%%%%%%%%%%%%%%%%%%%%%%%%%%%%%%%%%%%
%                      PAGE DE TITRE
%%%%%%%%%%%%%%%%%%%%%%%%%%%%%%%%%%%%%%%%%%%%%%%%%%%%%%%%%%%%%%%%%%%%%%
\begin{titlepage}
April 1998\hfill PAR-LPTHE 98/12
\vskip 5cm
{\baselineskip 17pt
\begin{center}
{\bf INDIRECT \mathversion{bold}$CP$ VIOLATION IN AN ELECTROWEAK
\noindent\mathversion{bold}$SU(2)_L \times U(1)$ GAUGE THEORY
OF CHIRAL MESONS.}
\end{center}
}
\vskip .5cm
\centerline{B. Machet
     \footnote[1]{Member of `Centre National de la Recherche Scientifique'.}
     \footnote[2]{E-mail: machet@lpthe.jussieu.fr.}
           }
\vskip 5mm
\centerline{{\em Laboratoire de Physique Th\'eorique et Hautes Energies,}
     \footnote[3]{LPTHE tour 16\,/\,1$^{er}\!$ \'etage,
          Universit\'e P. et M. Curie, BP 126, 4 place Jussieu,
          F 75252 PARIS CEDEX 05 (France).}
}
\centerline{\em Universit\'es Pierre et Marie Curie (Paris 6) et Denis
Diderot (Paris 7);} \centerline{\em Unit\'e associ\'ee au CNRS UMR 7589.}
\vskip 1.5cm
%%%%%%%%%%%%%%%%%%%%%%%%%%%%%%%%%%%%%%%%%%%%%%%%%%%%%%%%%%%%%%%%%%%%%%%%%%
%                          ABSTRACT
%%%%%%%%%%%%%%%%%%%%%%%%%%%%%%%%%%%%%%%%%%%%%%%%%%%%%%%%%%%%%%%%%%%%%%%%%%
{\bf Abstract:}  Indirect $CP$ violation is analyzed in the framework of the
electroweak  gauge theory of $J=0$ mesons proposed in \cite{Machet1}
in which they transform
like composite fermion-antifermion operators by the chiral $U(N)_L \times
U(N)_R$ group and by the $SU(2)_L \times U(1)$ gauge group of the
Glashow-Salam-Weinberg model \cite{GlashowSalamWeinberg}.
It is shown that, in this model where, in particular, mass terms are
introduced for the mesons themselves, and  unlike what happens in the
standard model for
fermions \cite{CabibboKobayashiMaskawa}:\l
- electroweak mass eigenstates can differ from $CP$ eigenstates even in the
case of two generations;\l
- the existence of a complex entry in the mixing matrix
for the constituent fermions is no longer a sufficient condition for
indirect $CP$ violation to occur at the mesonic level.
\smallskip

% \centerline{\rule {3cm} {0.2pt}}
\vfill
\null\hfil\epsffile{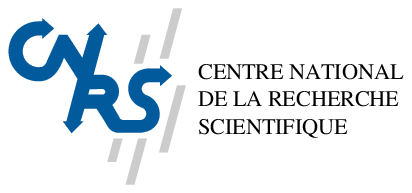}
\end{titlepage}
%%%%%%%%%%%%%%%%%%%%%%%%%%%%%%%%%%%%%%%%%%%%%%%%%%%%%%%%%%%%%%%%%%%%%%%%%%%
%                           TEXTE DU PAPIER
%%%%%%%%%%%%%%%%%%%%%%%%%%%%%%%%%%%%%%%%%%%%%%%%%%%%%%%%%%%%%%%%%%%%%%%%%%%
\section{Introduction; theoretical setting.}
\label{section:intro}
%%%%%%%%%%%%%%%%%%%%%%%%%%%%%%%%%%%%%%%%%%%%%%%%%%%%%%%%%%%%%%%%%%%%%%
%{\enlumin{T}}
%

The interpretation of mesons as fermion-antifermion composites \cite{GellMann}
is widely accepted. However, a field theory in which the fields in the
Lagrangian (quarks) are not the particles (asymptotic states)  steps on the
unsolved problem of confinement \cite{QCD}.
To bypass this difficulty, I proposed in
\cite{Machet1} a gauge theory in which $J=0$ mesons are both the fields and
the particles, but in which they transform, by the relevant symmetry groups,
like $\bar q_i q_j$ or $\bar q_i \gamma_5 q_j$ operators. It incorporates
in particular the chiral properties of the quarks $q$.
Requiring that, in the quest for unification, the gauge group of symmetry be
the same as for leptons, it is chosen to be the $SU(2)_L \times U(1)$ gauge
group of the Glashow-Salam-Weinberg model \cite{GlashowSalamWeinberg} and acts
on the ``constituent'' fermions accordingly; so, the symmetry properties of
the mesons also reflect the underlying electroweak symmetry of the standard
model of quarks. The latter are however no longer considered as dynamical
objects (they do not appear in the Lagrangian).

Let 
\begin{equation}
\Psi =
\left(
\ba{c}  u\\ c\\ \vdots \\d\\ s\\ \vdots \ea
\right)
\label{eq:Psi}\end{equation}
be a $N$-vector of fermions lying in the fundamental representation of $U(N)$.
There are $N/2$ families of fermions ($N$ is restricted to be even).

Any meson of the type $\bar q_i q_j$ or $\bar q_i \gamma_5 q_j$ is
represented by $\ol \Psi {\mathbb M} \Psi$ or
$\ol \Psi \gamma_5{\mathbb M} \Psi$, where $\mathbb M$ is an $N\times N$ matrix
with a single nonvanishing entry, equal to $1$, at the crossing of the
$i^{th}$ line and $j^{th}$ column; such matrices define the ``flavour'' or
``strong'' eigenstates (the flavour group of symmetry is the diagonal subgroup
of the chiral group, and supposed unbroken by strong interactions).
Any $N \times N$ matrix $\mathbb M$ represents, up
to the parity quantum number, a $J=0$ meson which is a linear combination of the
previous eigenstates; its behaviour, when acted upon by a symmetry group
is determined by the laws of transformations of the fermions.
The $\mathbb M$'s are the dynamical meson fields of the model.

The $\gamma_5$ matrix plays an essential role in the transformation of the
composite pseudoscalar operators by the chiral group;
it in particular introduces, in addition to commutators, anticommutators
of $\mathbb M$ with the generators of $U(N)_L \times U(N)_R$, which are
$N\times N$ matrices, too. The action of the chiral group on
the mesons consequently involves the associative character of the $U(N)$
algebra \cite{Machet1}.

Since we want to drop any reference to fermions, hence to $\gamma_5$, we
swap the latter for a doubling in the space of $\mathbb M$ matrices and
we distinguish ${\mathbb M}_{{\cal P}even}$ and
${\mathbb M}_{{\cal P}odd}$ mesons, corresponding to the combinations
$\ol \Psi {\mathbb M} \Psi + \ol \Psi \gamma_5{\mathbb M} \Psi$ and
$\ol \Psi {\mathbb M} \Psi - \ol \Psi \gamma_5{\mathbb M} \Psi$, which are
respectively 
even or odd by the action of the parity changing operator $\cal P$, and in
terms of which (see \cite{Machet1}) the laws of transformation by the
chiral group $U(N)_L \times U(N)_R$ are specially simple. The $2N^2$
independent $\mathbb{M}$ matrices so obtained match the total number
of $J=0$ scalar and pseudoscalar mesons built with $N$ flavours of fermions.

Care has also to be taken of the role of $\gamma_5$ as far as the
transformation by charge conjugation is concerned (see below).

The electroweak gauge
group naturally appears as a subgroup of the chiral group.
The three $SU(2)_L$ generators $\vec\mathbb{T}_L$ are also now $N\times N$
matrices (see \cite{Machet1} and the introduction of section
\ref{section:CP} below)
\begin{equation}
{\mathbb T}^3_L = {1\over 2}\left(\begin{array}{rrr}
                        {\mathbb I} & \vline & 0\\
                        \hline
                        0 & \vline & -{\mathbb I}
\end{array}\right),\
{\mathbb T}^+_L =           \left(\begin{array}{ccc}
                        0 & \vline & {\mathbb K}\\
                        \hline
                        0 & \vline & 0           \end{array}\right),\
{\mathbb T}^-_L =           \left(\begin{array}{ccc}
                        0 & \vline & 0\\
                        \hline
                        {\mathbb K}^\dagger & \vline & 0
\end{array}\right),
\label{eq:SU2L}
\end{equation}
which act trivially on the $N$-vector $\Psi_L = ((1-\gamma_5)/2)\Psi$.
$\mathbb I$ and $\mathbb K$ are respectively the $N/2 \times N/2$ identity
matrix and the most general unitary mixing matrix, which can in particular
be of the Cabibbo-Kobayashi-Maskawa (CKM) type \cite{CabibboKobayashiMaskawa}.
Consequences of the relationship that results between the electroweak
and chiral breaking have been emphasized in \cite{Machet2}.

The $U(1)$ generator satisfies the Gell-Mann-Nishijima relation
\cite{GellMannNishijima} (written in its ``chiral'' form)
\begin{equation}
({\mathbb Y}_L,{\mathbb Y}_R) =
                       ({\mathbb Q}_L,{\mathbb Q}_R) - ({\mathbb T}^3_L,0),
\label{eq:GMN}\end{equation}
and that the electric charge be represented by the customary (diagonal) operator
\begin{equation}
{\mathbb Q}_L ={\mathbb Q}_R ={\mathbb Q}=\left(\begin{array}{ccc}
                        2/3 & \vline & 0\cr
                        \hline
                        0 & \vline & -1/3
           \end{array}\right),
\label{eq:Q}\end{equation}
yields back the usual expressions for the ``left'' and ``right'' hypercharges
\begin{equation}
{\mathbb Y}_L = {1\over 6}{\mathbb I}, \quad {\mathbb Y}_R = {\mathbb Q}_R.
\label{eq:Y}\end{equation}
$\mathbb Q$ turns out to be the ``third'' generator of the custodial $SU(2)_V$
symmetry investigated in \cite{Machet1,Machet2}.

The orientation of the electroweak subgroup inside the chiral group is
controlled by a unitary matrix, $({\mathbb R},{\mathbb R})$, acting diagonally:
\begin{equation}
{\mathbb R} =             \left(\begin{array}{ccc}
                        {\mathbb I} & \vline & 0\\
                        \hline
                        0 & \vline & {\mathbb K}           \end{array}\right);
\label{eq:rotation}
\end{equation}
indeed, the electroweak group defined by eq.~(\ref{eq:SU2L}) is the one with
generators
\begin{equation}
{\mathbb R}^\dagger \vec t_L\; {\mathbb R}\; ;
\label{eq:rotgroup}
\end{equation}
with
\begin{equation}
{t}^3_L = {1\over 2}\left(\begin{array}{ccc}
                        {\mathbb I} & \vline & 0\\
                        \hline
                        0 & \vline & -{\mathbb I}
\end{array}\right),\
{t}^+_L =           \left(\begin{array}{ccc}
                        0 & \vline & {\mathbb I}\\
                        \hline
                        0 & \vline & 0           \end{array}\right),\
{t}^-_L =           \left(\begin{array}{ccc}
                        0 & \vline & 0\\
                        \hline
                        {\mathbb I} & \vline & 0           \end{array}\right).
\label{eq:generic}
\end{equation}
In practice, this rotation only acts on the ${t}^\pm$ generators
(we require ${t}^- = ({t}^+)^\dagger$, such that the
unit matrices in eqs.~(\ref{eq:generic},\ref{eq:SU2L}) have the same
dimension).

The $2N^2$ electroweak eigenstates can be classified into $N^2/2$ quadruplets,
split into two sets, respectively ``even'' and ``odd'' by the parity
changing operator $\cal P$.  All of them can be written in the form
\cite{Machet1}

\vbox{
\bea
& &\Phi(\mathbb D)=
({\mathbb M}\,^0, {\mathbb M}^3, {\mathbb M}^+, {\mathbb M}^-)(\mathbb D)\cr
& &\ \cr
& & =\left[
 {1\over \sqrt{2}}\left(\begin{array}{ccc}
                     {\mathbb D} & \vline & 0\\
                     \hline
                     0 & \vline & {\mathbb K}^\dagger\,{\mathbb D}\,{\mathbb K}
                   \end{array}\right),
{i\over \sqrt{2}} \left(\begin{array}{ccc}
                     {\mathbb D} & \vline & 0\\
                     \hline
                     0 & \vline & -{\mathbb K}^\dagger\,{\mathbb D}\,{\mathbb K}
                   \end{array}\right),
i\left(\begin{array}{ccc}
                     0 & \vline & {\mathbb D}\,{\mathbb K}\\
                     \hline
                     0 & \vline & 0           \end{array}\right),
i\left(\begin{array}{ccc}
                     0 & \vline & 0\\
                     \hline
                     {\mathbb K}^\dagger\,{\mathbb D} & \vline & 0
                    \end{array}\right)
             \right],\cr
& &
\label{eq:reps}
\eea
}
where $\mathbb D$ is a real $N/2 \times N/2$ matrix.

That the entries ${\mathbb M}^+$ and ${\mathbb M}^-$ are, up to a sign,
hermitian conjugate  requires that the
$\mathbb D$'s are restricted to symmetric or antisymmetric matrices.
Because of the presence of an ``$i$'' for the for ${\mathbb M}^{3,\pm}$ and
not for ${\mathbb M}^0$, the quadruplets (\ref{eq:reps})
always mix entries which have
different behaviour by hermitian  conjugation: they are consequently not
hermitian representations.

Each of them is  the sum of two
doublets of $SU(2)_L$, and also the sum of one singlet plus one triplet of
the custodial diagonal $SU(2)_V$ \cite{Machet1,Machet2}.
The $\cal P$-even and $\cal P$-odd
quadruplets do not transform in the same way
by $SU(2)_L$ (the Latin indices $i,j,k$ below run from $1$ to $3$);
for ${\cal P}$-even quadruplets, one has

\vbox{
\bea
{\mathbb T}^i_L\,.\,{\mathbb M}^j_{{\cal P}even} &=& -{i\over 2}\left(
              \epsilon_{ijk} {\mathbb M}^k_{{\cal P}even} +
                           \delta_{ij} {\mathbb M}_{{\cal P}even}^0
                              \right),\cr
{\mathbb T}^i_L\,.\,{\mathbb M}_{{\cal P}even}^0 &=&
                                {i\over 2}\; {\mathbb M}_{{\cal P}even}^i;
\label{eq:actioneven}
\eea
}
while ${\cal P}$-odd quadruplets transform according to

\vbox{
\bea
{\mathbb T}^i_L\,.\,{\mathbb M}_{{\cal P}odd}^j &=& -{i\over 2}\left(
                   \epsilon_{ijk} {\mathbb M}_{{\cal P}odd}^k -
                           \delta_{ij} {\mathbb M}_{{\cal P}odd}^0
                              \right),\cr
{\mathbb T}^i_L\,.\,{\mathbb M}_{{\cal P}odd}^0 &=&
                        \hskip 5mm  -{i\over 2}\; {\mathbb M}_{{\cal P}odd}^i,
\label{eq:actionodd}
\eea
}
and only representations transforming alike, $\cal P$-even or $\cal P$-odd,
can be linearly mixed.
The (diagonal) charge operator acts indifferently on both types of
representations by:

\vbox{
\bea
{\mathbb Q}\,.\,{\mathbb M}^i &=& -i\,\epsilon_{ij3} {\mathbb M}^j,\cr
{\mathbb Q}\,.\,{\mathbb M}^0 &=& 0.
\label{eq:Qaction}
\eea
}
The misalignment of ``strong'' and electroweak eigenstates, resulting from
the one of the electroweak group inside the chiral group, is conspicuous
from the presence of the mixing matrix $\mathbb K$ in (\ref{eq:reps}).

Adding or subtracting eqs.~(\ref{eq:actioneven}) and (\ref{eq:actionodd}),
and defining scalar ($\mathbb S$) and pseudoscalar ($\mathbb P$) fields by
\begin{equation}
({\mathbb M}_{{\cal P}even} + {\mathbb M}_{{\cal P}odd}) = {\mathbb S},
\label{eq:scalar}
\end{equation}
and
\begin{equation}
({\mathbb M}_{{\cal P}even} - {\mathbb M}_{{\cal P}odd}) = {\mathbb P},
\label{eq:pseudo}
\end{equation}
yields two new types of stable quadruplets which include objects of different
but definite parities, corresponding to  $CP$ eigenstates, depending
whether $\mathbb D$ is a symmetric or antisymmetric matrix
\begin{equation}
\varphi = ({\mathbb S}^0, \vec {\mathbb P}),
\label{eq:SP}
\end{equation}
and
\begin{equation}
\chi = ({\mathbb P}\,^0, \vec {\mathbb S});
\label{eq:PS}
\end{equation}
they transform in the same way by the gauge group, according to
eq.~(\ref{eq:actioneven}), and thus can be linearly mixed.  As they span the
whole space of $J=0$ mesons too, this last property makes them specially
convenient to consider.

Taking the hermitian conjugate of any representation $\Phi$ swaps the
relative sign between ${\mathbb M}^0$ and $\vec{\mathbb M}$; as a consequence,
$\Phi^\dagger_{{\cal P}even}$ transforms by $SU(2)_L$
as would formally do a ${\cal P}$-odd representation, and vice-versa;
on the other hand, the quadruplets (\ref{eq:reps}) are also representations
of $SU(2)_R$, the action of which is obtained by swapping
eqs.~(\ref{eq:actioneven}) and (\ref{eq:actionodd}) \cite{Machet1,Machet2};
so, the hermitian conjugate of a given representation of $SU(2)_L$ is a
representation of $SU(2)_R$ with the same law of transformation, and
vice-versa. The same result holds for any (complex) linear
representation  of quadruplets transforming alike by the gauge group.

The conjugate of a pseudoscalar $\bar q_i \gamma_5 q_j$ composite
operator being $- \bar q_j \gamma_5 q_i$ (the ``-'' sign comes from
the anticommutation of $\gamma_5$ with $\gamma^0$), one
must distinguish between the matrix conjugate ${\mathbb M}^\dagger$
and the charge conjugate $\ol{\mathbb M}$ of
any $\mathbb M$ meson: the latter, corresponding to charge conjugation,
gets an extra ``-'' sign with respect to the former when pseudoscalars
are concerned;
one has the following relation between the charge and hermitian conjugates
of parity eigenstates:
\begin{equation}
\ol {\mathbb S} = {\mathbb S}^\dagger, \quad
\ol {\mathbb P} = -{\mathbb P}^\dagger,
\label{eq:C1}\end{equation}
yielding in particular:\l
- for symmetric $\mathbb D$'s:
\begin{equation}
\ol {({\mathbb S}^0, \vec {\mathbb P})_{sym}}
                               = ({\mathbb S}^0, \vec {\mathbb P})_{sym},
\quad
\ol {({\mathbb P}^0, \vec {\mathbb S})_{sym}}
                               = -({\mathbb P}^0, \vec {\mathbb S})_{sym}.
\label{eq:C2}\end{equation}
- for antisymmetric $\mathbb D$'s:
\begin{equation}
\ol {({\mathbb S}^0, \vec {\mathbb P})_{antisym}}
                               = -({\mathbb S}^0, \vec {\mathbb P})_{antisym},
\quad
\ol {({\mathbb P}^0, \vec {\mathbb S})_{antisym}}
                               = ({\mathbb P}^0, \vec {\mathbb S})_{antisym}.
\label{eq:C3}\end{equation}
Only for $({\mathbb S}^0, \vec {\mathbb P})$ and $({\mathbb P}^0, \vec
{\mathbb S})$ quadruplets
is there any link between matrix and charge conjugation.
For $\Phi_{{\cal P}even}$ and $\Phi_{{\cal P}odd}$ quadruplets, which mix
scalars and pseudoscalars, one has instead the relations:
\begin{equation}
\ol {\Phi_{{\cal P}even}}= (\Phi_{{\cal P}odd})^\dagger,\quad
\ol {\Phi_{{\cal P}odd}}= (\Phi_{{\cal P}even})^\dagger.
\label{eq:C4}\end{equation}
The link between the dynamical (matricial) fields and the usually defined $J=0$
``strong'' mesonic eigenstates is the following: consider for example the
case $N=4$, for which the matrix $\mathbb K$
shrinks back to the Cabibbo matrix; the pseudoscalar $\pi^+$ meson, which
is a flavour or ``strong eigenstate, is represented in our notation,
up to a normalization factor, by the matrix-valued field
\begin{equation}
\Pi^+ = i\ \left( \ba{rrcrr}     &  &\vline & 1 &  0 \\
                                 &  &\vline & 0 &  0 \\
                            \hline
                                 &  &\vline &   &     \\
                                 &  &\vline &   &  \ea \right);
\label{eq:pi+}\end{equation}
sandwiched between two 4-vectors $\Psi$ of quarks (\ref{eq:Psi}),
it gives, indeed (restoring the $\gamma_5$), the wave function
\begin{equation}
\ol\Psi\ \gamma_5\Pi^+\ \Psi  = i\ \bar u \gamma_5 d,
\end{equation}
of the $(+1)$ charged pion \footnote{the $i$ factor ensures that $\ol{\pi^+}
= \pi^-$}.

The normalization of the state, determined from its leptonic decays
(see \cite{Machet3,Machet2,Machet1,Machet4}), is $a = 2f/\la H\ra$:
the relation between the  $\Pi^+$ matrix and the  field
(of dimension $[mass]$) of  the observed $\pi^+$ particle is $\Pi^+ = a \pi^+$.
$f$ is the leptonic decay constant of the mesons (considered to
be the same for all of them)  and $H = {\mathbb S}^0 ({\mathbb D}_1)$ is
the Higgs boson (see the remark at the end of Appendix A).

One likewise identifies the other  pseudoscalar mesons, $K^+, D^+, D_s^+$,
such that ${\mathbb P}^+({\mathbb D}_1)$
(see section \ref{section:invar} and Appendix A)
\begin{equation}
{\mathbb P}^+({\mathbb D}_1) =
i \left(\ba{rrcrr}   &  &\vline & c_\theta &  s_\theta \\
                             &  &\vline &-s_\theta &  c_\theta \\
                            \hline
                             &  &\vline &   &     \\
                             &  &\vline &   &  \ea \right),
\label{eq:P+}\end{equation}
one of the three Goldstone bosons of the broken electroweak symmetry,
is the linear combination of pseudoscalar mesons
\begin{equation}
{\mathbb P}^+({\mathbb D}_1) = {2f\over \la H\ra}
\left(c_\theta (\pi^+ + D_s^+) + s_\theta (K^+ -D^+)\right).
\label{eq:Goldstone}\end{equation}
%

%%%%%%%%%%%%%%%%%%%%%%%%%%%%%%%%%%%%%%%%%%%%%%%%%%%%%%%%%%%%%%%%%%%%%%
\section{The quadratic invariants.}
\label{section:invar}
%%%%%%%%%%%%%%%%%%%%%%%%%%%%%%%%%%%%%%%%%%%%%%%%%%%%%%%%%%%%%%%%%%%%%%

In order to construct a $SU(2)_L \times U(1)$ Lagrangian, one needs invariant
quadratic polynomials in the $J=0$ mesonic fields.

To every quadruplet $({\mathbb M}^0, \vec{\mathbb M})$
is associated such an invariant:
\begin{equation}
{\cal I}
= ({\mathbb M}^0, \vec {\mathbb M})\otimes ({\mathbb M}^0, \vec {\mathbb M})
= {\mathbb {\mathbb M}}\,^0 \otimes {\mathbb {\mathbb M}}\,^0 +
                 \vec {\mathbb M} \otimes \vec {\mathbb M};
\label{eq:invar}
\end{equation}
the ``$\otimes$'' product is a tensor product, not the usual multiplication
of matrices and means the product of fields as functions of space-time;
$\vec {\mathbb M} \otimes \vec {\mathbb M}$ stands for
$\sum_{i=1,2,3} {\mathbb M}\,^i \otimes  {\mathbb M}\,^i$.

For the relevant cases $N=2,4,6$, there exists a set of $\mathbb D$ matrices
(see appendix A) such that the algebraic sum 
of invariants specified below, extended over all  representations defined by
(\ref{eq:SP},\ref{eq:PS},\ref{eq:reps})
\bea
&&
{1\over 2}
\left((\sum_{symmetric\ {\mathbb D}'s} - \sum_{antisym\ {\mathbb D}'s})
\left( ({\mathbb S}^0, \vec {\mathbb P})({\mathbb D})
                     \otimes  ({\mathbb S}^0, \vec {\mathbb P})({\mathbb D})
- ({\mathbb P}^0, \vec {\mathbb S})({\mathbb D})
                     \otimes  ({\mathbb P}^0, \vec {\mathbb S})({\mathbb D})
\right)\right)\cr
&=&
{1\over 2}
\left(\sum_{all\ {\mathbb D}'s}
\left( \ol{({\mathbb S}^0, \vec {\mathbb P})({\mathbb D})}
                   \otimes  ({\mathbb S}^0, \vec {\mathbb P})({\mathbb D})
+ \ol{({\mathbb P}^0, \vec {\mathbb S})({\mathbb D})}
                   \otimes  ({\mathbb P}^0, \vec {\mathbb S})({\mathbb D})
\right)\right)\cr
&=&
{1\over 4}
\left(\sum_{all\ {\mathbb D}'s}
\left(\Phi^\dagger_{{\cal P}odd}({\mathbb D})
                        \otimes \Phi_{{\cal P}even}({\mathbb D})
+\Phi_{{\cal P}odd}({\mathbb D})
                        \otimes \Phi_{{\cal P}odd}({\mathbb D})
\right) \right)\cr
&=&
{1\over 4}
\left(\sum_{all\ {\mathbb D}'s}
\left(\ol{\Phi_{{\cal P}even}({\mathbb D})}
                            \otimes \Phi_{{\cal P}even}({\mathbb D})
+ \ol{\Phi_{{\cal P}odd}({\mathbb D})}
                             \otimes \Phi_{{\cal P}odd}({\mathbb D})
\right) \right)\cr
&&
\label{eq:Idiag}\eea
is diagonal both in the electroweak basis and in the basis of strong
eigenstates:
in the last one, all terms are normalized alike to $(+1)$.
Two ``$-$'' signs  occur in the first line of  eq.~(\ref{eq:Idiag})
\footnote{Eq.~(\ref{eq:Idiag} specifies eq.~(25) of
\cite{Machet1}, in which the ``$-$'' signs were not explicitly written.}
:\l
- the first between the $({\mathbb P}^0, \vec{\mathbb S})$ and
$({\mathbb S}^0, \vec{\mathbb P})$ quadruplets, because, as seen on
eq.~(\ref{eq:reps}), the ${\mathbb P}^0$ entry of the former has no ``$i$''
factor, while the $\vec{\mathbb P}$'s of the latter do have one; as we define
all pseudoscalars with an ``$i$'' (see
eqs.~(\ref{eq:pi+},\ref{eq:P+},\ref{eq:Goldstone})),
a $(\pm i)$ relative factor has to be introduced
between the two types of representations, yielding a ``$-$'' sign in
eq.~(\ref{eq:Idiag});\l
- the second for  the representations corresponding to antisymmetric
$\mathbb D$ matrices, which have an opposite behaviour by matrix conjugation 
as compared to the ones with symmetric ${\mathbb D}$'s.

The kinetic part of the $SU(2)_L \times U(1)$ Lagrangian for $J=0$ mesons
is built from the combination (\ref{eq:Idiag}) of invariants above, now used
for the covariant derivatives of the fields with respect to the gauge group;
it is thus  also diagonal in both the strong and electroweak basis.

Other invariants can be built which are tensor products of two representations
transforming alike by the gauge group: two $\cal P$-odd or two $\cal P$-even,
two $({\mathbb S}^0,\vec {\mathbb P})$'s,
two $({\mathbb P}^0,\vec {\mathbb S})$'s,
or one $({\mathbb S}^0,\vec {\mathbb P})$ and one
$({\mathbb P}^0,\vec {\mathbb S})$; such is for example 
\begin{equation}
{\cal I}_{1\ti 2} = ({\mathbb S}^0,\vec {\mathbb P})({\mathbb D}_1) \otimes
                ({\mathbb P}^0,\vec {\mathbb S})({\mathbb D}_2)
 ={\mathbb S}^0({\mathbb D}_1) \otimes {\mathbb P}^0({\mathbb D}_2) +
        \vec {\mathbb P}({\mathbb D}_1) \otimes \vec {\mathbb S}({\mathbb D}_2).
\label{eq:I12}\end{equation}
According to the remark made in the previous section, all the above
expressions are also invariant by the action of $SU(2)_R$.

The quadratic $SU(2)_L$ invariants (\ref{eq:invar}) are not
{\em a priori} self conjugate expressions and have consequently  no definite
property by hermitian conjugation.

However, unitarity compels the electroweak Lagrangian of $J=0$ mesons to be
hermitian.

Any invariant quadratic expression constructed from the quadruplets
(\ref{eq:SP},\ref{eq:PS}) is hermitian (a ``$i$'' factor has eventually to
be introduced), like in particular the expression (\ref{eq:Idiag})
that yields the kinetic terms when used for the covariant derivatives of
the fields; each entry of theirs has indeed a well defined behaviour by
hermitian conjugation.

In general, the invariant $\Omega \otimes \Omega$ attached to a given
quadruplet $\Omega$  is hermitian if and only if the condition
$\ol\Omega = \pm \Omega$
is satisfied; from eqs.~(\ref{eq:C2},\ref{eq:C3}),
it can only be a representation of
the type (\ref{eq:SP}) or (\ref{eq:PS}).

For a general quadruplet (\ref{eq:reps}), for example $\Phi_{{\cal P}even}$,
the invariant $\Phi_{{\cal P}even} \otimes \Phi_{{\cal P}even}$
is not hermitian, its charge conjugate being, by eq.~(\ref{eq:C4})
\begin{equation}
\ol{\Phi_{{\cal P}even} \otimes \Phi_{{\cal P}even}} =
\Phi_{{\cal P}odd}^\dagger \otimes \Phi_{{\cal P}odd}^\dagger \not =
\pm \Phi_{{\cal P}even} \otimes \Phi_{{\cal P}even}.
\end{equation}
Instead, for the representations (\ref{eq:reps}), the quadratic expressions

\vbox{
\bea
\ol{\Phi_{{\cal P}even}} \otimes \Phi_{{\cal P}even}
              &=& \Phi_{{\cal P}odd}^\dagger \otimes \Phi_{{\cal P}even},\cr
&& \cr
\ol{\Phi_{{\cal P}odd}} \otimes \Phi_{{\cal P}odd}
              &=& \Phi_{{\cal P}even}^\dagger \otimes \Phi_{{\cal P}odd},
\label{eq:C5}\eea
}
are hermitian and each one connects two quadruplets transforming alike by
the gauge group ($\Phi_{{\cal P}even}$ and $\Phi_{{\cal P}odd}^\dagger$,
$\Phi_{{\cal P}odd}$ and $\Phi_{{\cal P}even}^\dagger$)
(see section \ref{section:intro}).

If the $\Phi$'s in (\ref{eq:C5}) correspond to matrices $\mathbb D$ with
definite symmetry properties,  the above hermitian invariants do not depend
whether $\Phi$ is ${\cal P}odd$ or ${\cal P}even$, but only whether
$\mathbb D$ is symmetric or skew-symmetric (this determines the signs in
the second lines of the two equalities in eq.~(\ref{eq:mphi}) below,
though the third lines are formally identical). One has

\vbox{
\bea
{\cal I}_\Phi^{sym} &=& 
     \ol{\Phi_{{\cal P}even}({\mathbb D}_{sym})}\otimes
                            \Phi_{{\cal P}even}({\mathbb D}_{sym})=
     \ol{\Phi_{{\cal P}odd}({\mathbb D}_{sym})}\otimes
                          \Phi_{{\cal P}odd}({\mathbb D}_{sym})\cr
&& \cr
&=&
      ({\mathbb S}^0, \vec {\mathbb P})({\mathbb D}_{sym})^{\otimes 2}
     -({\mathbb P}^0, \vec {\mathbb S})({\mathbb D}_{sym})^{\otimes 2}\cr
&& \cr
&=& \ol{({\mathbb S}^0, \vec {\mathbb P})({\mathbb D}_{sym})}\otimes
              ({\mathbb S}^0, \vec {\mathbb P})({\mathbb D}_{sym})
   + \ol{({\mathbb P}^0, \vec {\mathbb S})({\mathbb D}_{sym})}\otimes
           ({\mathbb P}^0, \vec {\mathbb S})({\mathbb D}_{sym}),\cr
&& \cr
{\cal I}_\Phi^{antisym} &=& 
     \ol{\Phi_{{\cal P}even}({\mathbb D}_{antisym})}\otimes
           \Phi_{{\cal P}even}({\mathbb D}_{antisym})=
     \ol{\Phi_{{\cal P}odd}({\mathbb D}_{antisym})}\otimes
             \Phi_{{\cal P}odd}({\mathbb D}_{antisym})\cr
&& \cr
&=&
     -({\mathbb S}^0, \vec {\mathbb P})({\mathbb D}_{antisym})^{\otimes 2}
     + ({\mathbb P}^0, \vec {\mathbb S})({\mathbb D}_{antisym})^{\otimes 2}\cr
&&  \cr
&=& \ol{({\mathbb S}^0, \vec {\mathbb P})({\mathbb D}_{antisym})}\otimes
        ({\mathbb S}^0, \vec {\mathbb P})({\mathbb D}_{antisym})
   + \ol{({\mathbb P}^0, \vec {\mathbb S})({\mathbb D}_{antisym})}\otimes
   ({\mathbb P}^0, \vec {\mathbb S})({\mathbb D}_{antisym});\cr
&&
\label{eq:mphi}\eea
}
they correspond, for a given $\mathbb D$ matrix,
to degenerate $({\mathbb S}^0, \vec {\mathbb P})(\mathbb D)$
and $({\mathbb P}^0, \vec {\mathbb S})(\mathbb D)$ quadruplets. 

The expressions above
\footnote{and other invariant tensor products mixing quadruplets with
different symmetry properties like $\Phi_{{\cal P}even}({\mathbb D}_{antisym})
\otimes \Phi_{{\cal P}even}({\mathbb D}_{sym}) + h.c.$, 
$\ol{\Phi_{{\cal P}even}({\mathbb
D}_{antisym})}\otimes \Phi_{{\cal P}even}({\mathbb D}_{sym}) + h.c.$ \ldots }
can be used to build the  mass terms of the Lagrangian.

There are {\em a priori} as many $(N^2/2)$ independent mass scales as there are
independent representations (quadruplets).
Preserving the electroweak and custodial symmetries, they share
with the leptonic case the same arbitrariness; the hierarchy of mesonic masses
has here the same status as, e.g., the one between the muon and the electron.

The breaking of $SU(2)_L \times U(1)$ down to $U(1)_{em}$ is compatible with
preserving the custodial $SU(2)_V$ symmetry \cite{Machet1};
the number of allowed mass doubles up to $N^2$ and a splitting can occur
inside each quadruplet between the triplet and the singlet of
$SU(2)_V$ (this allows in particular a scalar-pseudoscalar
splitting inside each quadruplet (\ref{eq:SP}) or (\ref{eq:PS})).
This is to be compared with the standard model in which the custodial
symmetry is broken by any mass splitting inside a doublet of
$SU(2)_L$ \cite{Sikivie}.

As the electroweak group is a subgroup of the chiral group, chiral and
electroweak symmetry breaking are connected. In particular,
introducing a vacuum expectation value for the Higgs boson is equivalent
(see Appendix A) to allowing for diagonal $\la\bar q_i q_i\ra$ condensates.

The traditional picture \cite{CurrentAlgebra} associates the $N^2$
pseudoscalar mesons (flavour or ``strong'' eigenstates)
with the $N^2$ (pseudo)-Goldstones generated by
the breaking of the chiral group $U(N)_L \times U(N)_R$ into its diagonal
flavour subgroups.  They become (perturbatively) massive by electroweak
radiative corrections \cite{Weinberg} and the (non-perturbative)
bulk of their masses is parameterized by introducing (soft) mass terms for
the quarks in the QCD Lagrangian \cite{QCD}.
This point of view is now modified since:\l
- as observed experimentally, none of the $2N^2$ $J=0$ ``strong'' mesons
corresponds to a Goldstone
particle: in addition to electroweak radiative corrections, they can be given
arbitrary masses, the number of which depends on the symmetries that are
preserved (chiral, or electroweak, or just the custodial symmetry);\l
- scalar and pseudoscalar electroweak mass eigenstates are on the same
footing, but can be splitted while preserving the custodial symmetry;\l
- there exist only three genuine Goldstones bosons $\vec{\mathbb P}({\mathbb
D}_1)$ which arise when the electroweak symmetry is spontaneously broken
down to $U(1)_{em}$, or, equivalently, chiral $SU(2)_L \times SU(2)_R$ broken
into $SU(2)_V$; they are linear combinations of flavour
eigenstates (pseudoscalar mesons) (see eq.~(\ref{eq:Goldstone})).

The mass splittings, which can be arbitrarily large, have a purely electroweak
origin since, from  eq.~(\ref{eq:Idiag}), equal electroweak mass
terms also correspond to equal mass terms for strong eigenstates
\footnote{in particular, chiral symmetry can be preserved when all scalars
and pseudoscalars are degenerate in mass.}
; however, their hierarchy obviously lies outside the realm of perturbation
theory.

We define electroweak mass eigenstates as states which diagonalize
both the kinetic terms and the mass terms of the electroweak Lagrangian.

%%%%%%%%%%%%%%%%%%%%%%%%%%%%%%%%%%%%%%%%%%%%%%%%%%%%%%%%%%%%%%%%%%%%%%
\section{Indirect $\mathbf{CP}$ violation.}
\label{section:CP}
%%%%%%%%%%%%%%%%%%%%%%%%%%%%%%%%%%%%%%%%%%%%%%%%%%%%%%%%%%%%%%%%%%%%%%

If the stakes are high for the observation of ``direct'' $CP$ violation
\cite{Jarlskog,Nir}, and, in particular, for discovering whether the so-called
$\epsilon '$ parameter is vanishing or not \cite{NA48-KTEV},
all phenomena of $CP$ violation
\cite{ChristensonCroninFitchTurlay} \cite{Argus} are, up to now, compatible
with the so-called ``indirect'' violation \cite{Jarlskog,Nir},
which means that the electroweak mass eigenstates are not $CP$ eigenstates.

The only known mechanism to trigger it in the
electroweak standard model for fermions is through a complex mixing matrix
for quarks \cite{CabibboKobayashiMaskawa}
\footnote{Alternatives need enlarging the scalar sector of the model
\cite{LeeWeinberg}.};
for a number of generations at least equal to three, in the process of
diagonalization of the most general mass matrix for quarks, one phase
cannot be reabsorbed in their wave functions. Hence, the combined experimental
evidence for indirect $CP$ violation and the preeminence of the
Glashow-Salam-Weinberg model as ``the'' theoretical tool to interpret the
data led, before its discovery \cite{bottom,top}, to a strong  prejudice in
favour of the existence of a third generation of fermions.

Despite our conservative attitude, $CP$ violation nevertheless
appears now from a different perspective, which shows in particular how
sensitive can be the interpretation of experimental data to the theoretical
filter used for their analysis. The mechanism which was sufficient to
trigger it at the quark level, and which used to be extrapolated at the level
of the asymptotic states despite our ignorance
of the dynamics of confinement, now fails when the mesons themselves are
chosen to be the dynamical fields of the theory, though  the chiral
and electroweak symmetry properties of their building blocks
have been preserved. As mass terms are now introduced for the mesons
themselves and no longer for quarks, that all phases in the mixing matrix
can be or not reabsorbed has lost its previous relevance.

I show below that, in the $SU(2)_L \times U(1)$ electroweak gauge theory for
mesons (and leptons -- but in this sector it is the genuine
Glashow-Salam-Weinberg model --) proposed above, the role of the
mixing matrix fades away as far as indirect $CP$ violation is concerned.
It only now determines the orientation of the electroweak group inside the
chiral group; that it is the same matrix  as the one controlling, in the
standard electroweak  model, the misalignment between leptons and quarks, 
allows the identification (\ref{eq:SU2L}) of $SU(2)_L \times U(1)$ as
a precise subgroup of the chiral group. The identification that we make
here of the building blocks of the $\mathbb M$ mesons with the quarks of
the standard model  could eventually be loosened into  a less conservative
approach and more exotic enlargements of the latter.

Another mixing matrix now triggers indirect $CP$ violation, the one
 which diagonalizes the kinetic terms with a new set of states
different from the ``strong'' or ``flavour'' eigenstates (\ref{eq:SP},
\ref{eq:PS}) occurring in eq.~(\ref{eq:Idiag}); 
that it includes at least one complex entry is  the necessary condition
for the existence of electroweak mass eigenstates for mesons which are not
$CP$ eigenstates.

%%%%%%%%%%%%%%%%%%%%%%%%%%%%%%%%%%%%%%%%%%%%%%%%%%%%%%%%%%%%%%%%%%%%%%
\subsection{Electroweak mass eigenstates can be $\mathbf{CP}$ eigenstates 
in the presence of a complex CKM matrix.}
\label{subsec:res1}

The demonstration of this first result is straightforward.

In the following Lagrangian for $J=0$ mesons, which is hermitian
and $SU(2)_L \times U(1)$ invariant, and  where the sum is extended to the
representations defined by eqs.~(\ref{eq:SP},\ref{eq:PS}) and all
$\mathbb D$ matrices defined in Appendix A:
($D_\mu$ is the covariant derivative with respect to $SU(2)_L \times U(1)$)

\vbox{
\bea
{\cal L}= &&{1\over 2}\sum_{symmetric\ {\mathbb D}'s}
           \left(D_\mu ({\mathbb S}^0, \vec {\mathbb P})(\mathbb D)
                  \otimes D^\mu ({\mathbb S}^0, \vec {\mathbb P})(\mathbb D)
          - m_D^2 ({\mathbb S}^0, \vec {\mathbb P})(\mathbb D)
                  \otimes ({\mathbb S}^0, \vec {\mathbb P})(\mathbb D)
        \right.\cr
&&\hphantom{{1\over 2}\sum_{symmetric\ {\mathbb D}}}
\left. -\left(D_\mu ({\mathbb P}^0, \vec {\mathbb S})(\mathbb D)
                    \otimes D^\mu ({\mathbb P}^0, \vec {\mathbb S})(\mathbb D)
          - \ti m_D^2 ({\mathbb P}^0, \vec {\mathbb S})(\mathbb D)
                    \otimes ({\mathbb P}^0, \vec {\mathbb S})(\mathbb D)
\right) \right)\cr
&& -\hskip 5mm{1\over 2}\sum_{antisym\ {\mathbb D}'s}
\left(D_\mu ({\mathbb S}^0, \vec {\mathbb P})(\mathbb D)
                     \otimes D^\mu ({\mathbb S}^0, \vec {\mathbb P})(\mathbb D)
          - m_D^2 ({\mathbb S}^0, \vec {\mathbb P})(\mathbb D)
                \otimes ({\mathbb S}^0, \vec {\mathbb P})(\mathbb D) \right.\cr
&&\hphantom{{1\over 2}\sum_{symmetric\ {\mathbb D}}}
        \left.-\left(D_\mu ({\mathbb P}^0, \vec {\mathbb S})(\mathbb D)
                     \otimes D^\mu ({\mathbb P}^0, \vec {\mathbb S})(\mathbb D)
          - \ti m_D^2 ({\mathbb P}^0, \vec {\mathbb S})(\mathbb D)
         \otimes ({\mathbb P}^0, \vec {\mathbb S})(\mathbb D)\right)\right),\cr
= && {1\over 2}\sum_{all\ {\mathbb D}'s}
           \left(D_\mu\ol{ ({\mathbb S}^0, \vec {\mathbb P})(\mathbb D)}
                \otimes D^\mu ({\mathbb S}^0, \vec {\mathbb P})(\mathbb D)
          - m_D^2 \ol{({\mathbb S}^0, \vec {\mathbb P})(\mathbb D)}
                    \otimes ({\mathbb S}^0, \vec {\mathbb P})(\mathbb D)
        \right.\cr
&&\hphantom{{1\over 2}\sum_{all\ {\mathbb D}}}
\left. +\left(D_\mu\ol{ ({\mathbb P}^0, \vec {\mathbb S})(\mathbb D)}
              \otimes D^\mu ({\mathbb P}^0, \vec {\mathbb S})(\mathbb D)
          - \ti m_D^2 \ol{({\mathbb P}^0, \vec {\mathbb S})(\mathbb D)}
               \otimes ({\mathbb P}^0, \vec {\mathbb S})(\mathbb D)
\right) \right),\cr
& &
\label{eq:LCP}\eea
}
the mass eigenstates, being the ${\mathbb S}^0, \vec {\mathbb P},
{\mathbb P}^0, \vec {\mathbb S}$ mesons, are also $CP$ eigenstates.
The complex matrix $\mathbb K$ is entirely absorbed in their definition,
and no complex coupling constant appears in the Lagrangian.

It is of course straightforward to build hermitian $SU(2)_L \times
U(1)$ invariant quartic terms.

Conclusion: {\em The existence of a complex phase in the mixing matrix for
quarks is not a sufficient condition for electroweak mass eigenstates of $J=0$
mesons transforming like $\bar q_i q_j$ or $\bar q_i \gamma_5 q_j$ to differ
from $CP$ eigenstates}.

%%%%%%%%%%%%%%%%%%%%%%%%%%%%%%%%%%%%%%%%%%%%%%%%%%%%%%%%%%%%%%%%%%%%%%
\subsection{Indirect $\mathbf{CP}$ violation can occur at the mesonic level
with only two generations of fermions.}
\label{subsec:res2}

Whatever be the number of generations, to every $\Phi_{{\cal P}even}$ or
$\Phi_{{\cal P}odd}$ quadruplet can be attached (see section
\ref{section:invar}) an $SU(2)_L \times U(1)$ invariant and hermitian
mass term $\ol\Phi \Phi$; hence it is trivial to have electroweak
mass eigenstates which are not $CP$ eigenstates because they are not
$P$ eigenstates. For example, any set of linear combinations of the
$\Phi_{{\cal P}even}({\mathbb D}_i)$ (see Appendix A)
that also diagonalize the kinetic terms can be mass eigenstates
corresponding to $(\mathbb{S +P})$ combinations.

However, it seems not to be questioned at present that observed
electroweak mass eigenstates have a well defined parity, and that,
consequently, indirect $CP$ violation can only spring from $C$ violation.

I show how, in the proposed model, electroweak
eigenstates of $J=0$ mesons can be $P$ but not $CP$ eigenstates with only
two generations of fermions.

The  simple mechanism lies in that the kinetic terms of the Lagrangian
can also be diagonalized with eigenvectors which are complex linear
combinations $\xi$ of the quadruplets of parity eigenstates
$({\mathbb S}^0, \vec {\mathbb P})$
(the same can be done with $({\mathbb P}^0, \vec {\mathbb S})$ quadruplets),
for which hermitian and $SU(2)_L \times U(1)$ invariant mass terms
$\ol\xi \xi$ can straightforwardly be written.

Let us work in the basis made with the quadruplets $\varphi$ and $\chi$
given by eqs.~(\ref{eq:SP},\ref{eq:PS}),
which we split into $\varphi_{sym}$, $\varphi_{antisym}$, $\chi_{sym}$,
$\chi_{antisym}$ according to the symmetry property of the
matrix $\mathbb D$.
We cast them into a vector $\Upsilon$ with dimension $2N^2$, written in
an abbreviated notation
\begin{equation}
\Upsilon =
\left(
\ba{c} \varphi_{sym}({\mathbb D})\\ \varphi_{antisym}({\mathbb D})\\
       \chi_{sym}({\mathbb D})\\ \chi_{antisym}({\mathbb D})\ea
\right).
\label{eq:Upsilon}\end{equation}
For example, for two generations ($N=4$), $\varphi_{sym}({\mathbb D})$ stands
above for
the three independent $({\mathbb S}^0, \vec {\mathbb P})({\mathbb D})$
quadruplets corresponding to the three $2\times 2$ symmetric
${\mathbb D}_1,{\mathbb D}_2,{\mathbb D}_3$ matrices (see Appendix A),
$\varphi_{antisym}({\mathbb D})$ for the unique
$({\mathbb S}^0, \vec {\mathbb P})({\mathbb D}_4)$ quadruplet corresponding to
the unique antisymmetric ${\mathbb D}_4$ matrix etc.

The kinetic terms of the Lagrangian write
\begin{equation}
{\cal L}_{kin}=
       \frac{1}{2}\ \p_\mu \ol{\Upsilon}\ [{\mathbb I}]\  \p^\mu \Upsilon
+\cdots
\label{eq:Lkin}\end{equation}
where $[{\mathbb I}]$ is the $2N^2 \times 2N^2$ unit matrix.

Let us make a change of basis described by the $2N^2 \times 2N^2$ matrix
$\mathbb U$
\begin{equation}
\Upsilon = {\mathbb U}\ \Xi,
\label{eq:change}\end{equation}
such that ${\mathbb U}^\dagger\  {\mathbb U}$ is diagonal.

$\Xi$ is a vector of dimension $2N^2$
\begin{equation}
\Xi = \left(
\ba{c} [\xi] \cr
       [\omega]
      \ea\right),
\label{eq:Xi}\end{equation}
in which the first $N^2$ entries,
generically called $[\xi]$ are linear combinations of the $\varphi(\mathbb
D)$'s,
and the last $N^2$, $[\omega]$ are linear combinations of the
$\chi(\mathbb D)$'s.

We furthermore request that
${\mathbb U}$ does not mix states with different parities: it writes
\begin{equation}
{\mathbb U} = \left( \ba{ccc} {\mathbb A} & \vline & 0 \cr
                       \hline
                      0 & \vline & {\mathbb B}
               \ea \right)
\label{eq:U}\end{equation}
where both $\mathbb A$ and $\mathbb B$ are $N^2 \times N^2$ matrices.
 
For the sake of simplicity, and without losing any generality, let us
suppose that ${\mathbb B}$ is the unit matrix, such that the new
eigenvectors diagonalizing the kinetic terms only differ from the original
ones in the subspace of $({\mathbb S}^0, \vec{\mathbb P})$ quadruplets.

The hermitian mass term that can be constructed for the $\xi$'s,
\begin{equation}
{\cal L}_m^\xi \propto \sum_{i=1\ldots N^2}{\mu_i^2\  \ol \xi_i \otimes \xi_i} 
\label{eq:Lm}\end{equation}
is $SU(2)_L \times U(1)$ invariant since it only involves invariant tensor
products of pairs or $({\mathbb S}^0, \vec{\mathbb P})$ quadruplets
transforming alike by the gauge group (see section \ref{section:intro}).

$\bullet$ If ${\mathbb A}$ is purely real, the mass eigenstates split into $CP$
eigenstates:
consider indeed a real linear combination $\xi$ among those which
diagonalize the kinetic terms
\begin{equation}
\xi = \sum_{i=1\ldots N_{sym}} a_i\  \varphi^i_{sym} +
              \sum_{i=1\ldots N_{antisym}} b_i\  \varphi^i_{antisym}
\label{eq:real}\end{equation}
with $a_i, b_i$ real, and $N_{sym} + N_{antisym} = N^2$.
The corresponding hermitian mass term for $\xi$ is proportional to

\vbox{
\bea
\hskip -1cm \ol\xi\xi 
&=& \left( \sum_{i=1\ldots N_{sym}} a_i\  \varphi^i_{sym} -
              \sum_{i=1\ldots N_{antisym}} b_i\  \varphi^i_{antisym}\right)
\otimes
            \left( \sum_{i=1\ldots N_{sym}} a_i\  \varphi^i_{sym} +
            \sum_{i=1\ldots N_{antisym}} b_i\  \varphi^i_{antisym}\right)\cr
& & \hphantom{www}\cr
& & \hphantom{www}\cr
&=& 
\left( \sum_{i=1\ldots N_{sym}} a_i\  \varphi^i_{sym}\right)^{\otimes\,2}
-\left(\sum_{i=1\ldots N_{antisym}} b_i\  \varphi^i_{antisym}\right)^{\otimes\,2},
\label{eq:Lreal}\eea
(where we have used  eqs.~(\ref{eq:C2},\ref{eq:C3}) for the last equality)
such that the  electroweak mass eigenstate $\xi$ splits into two degenerate
$CP$ eigenstates
$\sum_{i=1\ldots N_{sym}} a_i\  \varphi^i_{sym}$ and
$\sum_{i=1\ldots N_{antisym}} b_i\  \varphi^i_{antisym}$.
}

$\bullet$ If there is to be any indirect $CP$ violation, it can thus only
occur through a
complex mixing matrix between mesons. That it can indeed happen is easily
demonstrated on a simple example.

Consider in eq.~(\ref{eq:U}), still for $N=4$, a matrix $\mathbb A$
\begin{equation}
{\mathbb A} = \left( \ba{cccc}
                   1 & 0 & 0 & 0 \cr
                   0 & a & b & 0 \cr
                   0 & c & d & 0 \cr
                   0 & 0 & 0 & 1
      \ea \right).
\label{eq:A}\end{equation}
It couples here the two symmetric quadruplets
$\varphi_2 =({\mathbb S}^0, \vec {\mathbb P})({\mathbb D}_2)$ and
$\varphi_3 =({\mathbb S}^0, \vec {\mathbb P})({\mathbb D}_3)$
but  the demonstration can be made with any matrix $\mathbb A$.

That the kinetic term can be diagonalized with $\xi_2$ and $\xi_3$ as well
as with $\varphi_2$ and $\varphi_3$ requires
${\mathbb V}^\dagger {\mathbb V}$ to be diagonal,
${\mathbb V}$ being the $2\times 2$ complex matrix
\begin{equation}
{\mathbb V} = \left( \ba{cc} a & b \cr
                   c & d
    \ea \right);
\label{eq:V}\end{equation}
this gives the conditions
\begin{equation}
\bar a b + \bar c d = 0 = a \bar b + c \bar d.
\label{eq:diag}\end{equation}
Writing 
\begin{equation}
\Delta = ad - bc = -\frac{b}{\bar c}\ (\vert a \vert^2 + \vert c \vert^2),
\label{eq:Delta}\end{equation}
where we have  used (\ref{eq:diag}) for the last equality,
the new eigenstates  are

\vbox{
\bea
\xi_2 &=& \frac{1}{\Delta}(d \varphi_2 -b \varphi_3) = 
      -\frac{1}{\Delta}\ \frac{b}{\bar c}
                          \ (\bar a \varphi_2 + \bar c \varphi_3)\cr
&&\cr
\xi_3 &=& \frac{1}{\Delta}(-c \varphi_2 + a \varphi_3),
\label{eq:neweigen}\eea
}
for which one can introduce the hermitian mass terms with real coefficients

\vbox{
\bea
&& \mu_2^2\  \ol\xi_2 \xi_2 + \mu_3^2\  \ol\xi_3\xi_3 = \cr
&& \cr
&&
\frac{\mu_2^2}{\vert \Delta\vert^2}\left(
\vert d\vert^2 \varphi_2^{\otimes 2} +\vert b \vert^2 \varphi_3^{\otimes 2}
                    - (\bar b d +\bar d b)\varphi_2 \otimes \varphi_3\right)
+
\frac{\mu_3^2}{\vert \Delta\vert^2}\left(
\vert c\vert^2 \varphi_2^{\otimes 2} +\vert a \vert^2 \varphi_3^{\otimes 2}
                  - (\bar a c +\bar c a)\varphi_2 \otimes \varphi_3\right);
\cr
&&
\label{eq:Lmm}\eea
}
they are again $SU(2)_L \times U(1)$ invariant because they involve
tensor products of quadruplets $\varphi_2$ and $\varphi_3$
 behaving alike by the gauge group.

Take for example $a$ and $b$ real, $c$ complex;  the condition
(\ref{eq:diag}) entails that
$d = -(\bar a/\bar c)b = - abc/\vert c\vert^2$ is also complex.

The hermitian and $SU(2)_L \times U(1)$ invariant Lagrangian

\hskip -1cm\vbox{
\bea
{\cal L} &=& \frac{1}{2}(\p_\mu \ol{\varphi_2}\otimes\p^\mu \varphi_2 +
             \p_\mu \ol{\varphi_3}\otimes\p^\mu \varphi_3) + \cdots\cr
         &&-\frac{1}{2\vert \Delta\vert^2}\left(
 (\mu_2^2\vert d\vert^2 + \mu_3^2\vert c\vert^2)\varphi_2^{\otimes 2}
+(\mu_2^2\vert b \vert^2 +\mu_3^2\vert a \vert^2)\varphi_3^{\otimes 2}
-\left(\mu_2^2(\bar b d +\bar d b) + \mu_3^2(\bar a c +\bar c a)\right)
              \varphi_2 \otimes \varphi_3
\right) + \cdots\cr
         &\equiv& \frac{1}{2}
  (\vert a \vert^2 + \vert c \vert^2)
\ \left(\p_\mu \ol{\xi_2} \otimes\p^\mu \xi_2
             + \frac{\vert b \vert^2}{\vert c \vert^2}
                \ \p_\mu \ol{\xi_3} \otimes\p^\mu \xi_3 \right)+\cdots\cr
          &&-\frac{1}{2}(\mu_2^2\ \ol{\xi_2}\otimes\xi_2 +
                       \mu_3^2\ \ol{\xi_3}\otimes\xi_3) + \cdots\cr
&&
\label{eq:Linv}\eea
}

admits $\xi_2$ and $\xi_3$ as electroweak mass eigenstates; they are $P$
eigenstates, but {\em not} $CP$ eigenstates because they are not $C$
eigenstates.

Conclusion: {\em indirect $CP$ violation can occur with two generations
only for $J=0$ mesons transforming like $\bar q_i q_j$ or
$\bar q_i \gamma_5 q_j$ composite operators.}

Remark: one needs more than one generation to be able to combine several
quadruplets with the same definite parity quantum numbers.

%\newpage\null
%%%%%%%%%%%%%%%%%%%%%%%%%%%%%%%%%%%%%%%%%%%%%%%%%%%%%%%%%%%%%%%%%%%%%%
\section{Conclusion: outlook and perspectives.}
\label{section:conclusion}
%%%%%%%%%%%%%%%%%%%%%%%%%%%%%%%%%%%%%%%%%%%%%%%%%%%%%%%%%%%%%%%%%%%%%%

Renormalizable extensions of the electroweak standard model usually enlarge
its gauge group of symmetry \cite{GUTS}, eventually incorporate
a ``flavour'' or ``horizontal'' symmetry \cite{HORIZONTAL}, or supersymmetry
\cite{SUSY}, often increase the number of particles,
but seldom question the parallel between quarks and leptons.
Among other reasons, this attitude finds its justification
in the subtle mechanism of
cancelation of anomalies \cite{AdlerBellJackiw} between the two types of
fields \cite{BouchiatIliopoulosMeyer}. The price to pay lies in ad-hoc
procedures to circumvent the problem of confinement \cite{Quigg}.

Other extensions, based on effective Lagrangian \cite{CHIRALTHEORIES},
partially break the parallel mentioned above, incorporate constraints
imposed by chiral dynamics \cite{CurrentAlgebra} and some of the features
of Quantum Chromodynamics \cite{QCD}. but abandon renormalizability.

Some, still non-renormalizable, exploit the analogy between the high mass
limit of the Higgs boson \cite{LeeQuiggThacker} and $\sigma$-models
\cite{sigmamodel} to give predictions for a strongly interacting scalar
sector \cite{Gatto}.

Models with dynamical symmetry breaking are often plagued, too, with
non-renormalizability \cite{NambuJonaLasinio}, or with the difficult issue of
flavour changing neutral currents \cite{TECHNICOLOUR}

We followed here a different approach: preserving renormalizability and
limiting the spectrum to the one of observed particles, we chose
to break the parallel between quarks and leptons and to promote the symmetry
which exists between true asymptotic states, mesons (bosons) and leptons
(fermions).

The misalignment between the electroweak and the chiral groups of symmetry
reflects into the one between electroweak and flavour (or ``strong'')
eigenstates.
The first type of eigenstates being, unlike the second, {\em a priori}
linear combinations
of states of different parities and behaving differently by charge conjugation,
indirect $CP$ violation is naturally expected to occur. We have shown that
this is what happens, independently of the number of generations (if greater
than one), through a complex mixing matrix which now occurs at the mesonic
level.

The quadruplet $({\mathbb S}^0,\vec{\mathbb P})({\mathbb D}_1)$, isomorphic
to the complex scalar doublet of the Glashow-Salam-Weinberg model, includes
the Higgs boson and the three Goldstones of the broken electroweak symmetry.
The Goldstone triplet being directly related to observed pseudoscalar mesons
( and not to ``technimesons''), those turn out to be connected to the Higgs
boson by the electroweak group of symmetry.  The latter is consequently
expected to play a role in electroweak decays of pseudoscalar mesons.
This is the subject of a forthcoming work \cite{Machet5}.

The problem of the cancelation of anomalies now requires that the leptonic
sector be by itself anomaly-free, which happens, for example, if the observed
$V-A$ couplings are effective vertices of a more fundamental purely vectorial
theory (this has been investigated in \cite{BellonMachet}).

The question of higher spins \cite{Regge} for the mesons needs investigation,
and also the sector of baryons.  About the last point, it is tempting to
advocate for the existence of a
dual sector \cite{Olive} since charge quantization is effective as soon as
the custodial symmetry is preserved \cite{Machet1,Machet2} and dyon-like
solutions have been exhibited in a similar model \cite{Cho}. Then the
baryons could be thought of as extended objects (solitons) \cite{Skyrme}
which would be the strongly interacting fields of the dual sector.
This is currently under investigation.

\newpage\null
%%%%%%%%%%%%%%%%%%%%%%%%%%%%%%%%%%%%%%%%%%%%%%%%%%%%%%%%%%%%%%%%%%%%%%
{\Large\bf Appendix}
%%%%%%%%%%%%%%%%%%%%%%%%%%%%%%%%%%%%%%%%%%%%%%%%%%%%%%%%%%%%%%%%%%%%%%

\appendix

\section{Diagonalizing eq.~(\protect\ref{eq:Idiag}) in the basis of strong
eigenstates: a choice of $\mathbb D$ matrices.}

The property is most simply verified for the ``non-rotated''
$SU(2)_L \times U(1)$ group and representations corresponding to
eq.~(\ref{eq:generic}) and setting ${\mathbb K}={\mathbb I}$ in
(\ref{eq:reps}).

\subsection{$\mathbf N=2$ (1 generation).}

Trivial case: $\mathbb D$ is a number.

\subsection{$\mathbf N=4$ (2 generations).}

The four $2\times 2$ $\mathbb D$ matrices ($3$ symmetric and $1$
antisymmetric) can be taken as
\begin{equation}
{\mathbb D}_1 = \left( \ba{cc} 1 & 0 \cr
                            0 & 1     \ea \right),\
{\mathbb D}_2 = \left( \ba{rr} 1 & 0 \cr
                            0 & -1    \ea \right),\
{\mathbb D}_3 = \left( \ba{cc} 0 & 1 \cr
                            1 & 0     \ea \right),\
{\mathbb D}_4 = \left( \ba{rr} 0 & 1 \cr
                           -1 & 0     \ea \right).
\end{equation}

\subsection{$\mathbf N=6$ (3 generations).}

The nine $3 \times 3$ $\mathbb D$ matrices ($6$ symmetric and $3$
antisymmetric),
can be taken as
\bea
&&\hskip -1cm {\mathbb D}_1 = \sqrt{{2\over 3}}\left( \ba{ccc}
                                1  &  0  &  0  \cr
                                0  &  1  &  0  \cr
                                0  &  0  &  1 \ea \right), \cr
&& \cr
&& \cr
&&\hskip -1cm {\mathbb D}_2 ={2\over\sqrt{3}} \left( \ba{ccc}
                \sin\alpha  &     0    &    0    \cr
       0     & \sin(\alpha\pm{2\pi\over 3})&   0   \cr
       0     &                    0        & \sin(\alpha\mp{2\pi\over 3})
       \ea\right),\
{\mathbb D}_3 ={2\over\sqrt{3}} \left( \ba{ccc}
                \cos\alpha  &     0    &    0    \cr
       0     & \cos(\alpha\pm{2\pi\over 3})&   0   \cr
       0     &                    0        & \cos(\alpha\mp{2\pi\over 3})
       \ea\right), \cr
&& \cr
&& \cr
&&\hskip -1cm  {\mathbb D}_4 =\left( \ba{ccc}
                                0  &  0  &  1 \cr
                                0  &  0  &  0 \cr
                                1  &  0  &  0   \ea \right),\
     {\mathbb D}_5 =\left( \ba{rrr}
                                0  &  0  &  1 \cr
                                0  &  0  &  0 \cr
                               -1  &  0  &  0   \ea \right),\cr
&& \cr
&& \cr
&&\hskip -1cm {\mathbb D}_6 = \left( \ba{ccc}
                                0  &  1  &  0  \cr
                                1  &  0  &  0  \cr
                                0  &  0  &  0   \ea \right), \
 {\mathbb D}_7 = \left( \ba{rrr}
                                0  &  1  &  0  \cr
                               -1  &  0  &  0  \cr
                                0  &  0  &  0   \ea \right), \
 {\mathbb D}_8 = \left( \ba{ccc}
                                0  &  0  &  0  \cr
                                0  &  0  &  1  \cr
                                0  &  1  &  0   \ea \right), \
 {\mathbb D}_9 = \left( \ba{rrr}
                                0  &  0  &  0  \cr
                                0  &  0  &  1  \cr
                                0  & -1  &  0   \ea \right),\cr
& &
\eea
where $\alpha$ is an arbitrary  phase.

{\em Remark}: as ${\mathbb D}_1$ is the only matrix with a non vanishing trace,
${\mathbb S}^0({\mathbb D}_1)$ is the only neutral scalar matrix with the
same property;  we take it as the Higgs boson.

Considering that it is the only scalar with a non-vanishing vacuum
expectation value prevents the occurrence of a hierarchy problem
\cite{GildenerWeinberg}.

This last property is tantamount, in the ``quark language'', to taking
the same value for all condensates $\la\bar q_i q_i\ra, i=1 \cdots N$,
in agreement with the flavour independence of ``strong interactions''
between fermions, supposedly at the origin of this phenomenon in the
traditional framework.

As the spectrum of mesons is, in the present model, disconnected from a
hierarchy between quark condensates (see section \ref{section:invar}),
it is not affected by our choice of a single Higgs boson.
\newpage\null
%%%%%%%%%%%%%%%%%%%%%%%%%%%%%%%%%%%%%%%%%%%%%%%%%%%%%%%%%%%%%%%%%%%%%%%%
%                      REFERENCES
%%%%%%%%%%%%%%%%%%%%%%%%%%%%%%%%%%%%%%%%%%%%%%%%%%%%%%%%%%%%%%%%%%%%%%%%
\begin{em}

\end{em}


\begin{thebibliography}{50}
%
\bibitem{Machet1}
      B. MACHET: ``Chiral Scalar Fields, Custodial Symmetry in
                   Electroweak $SU(2)_L \times U(1)$, and the Quantization
                   of the Electric Charge'',
                   hep-ph/9606239,  Phys. Lett. B 385 (1996) 198-208.

\bibitem{GlashowSalamWeinberg}
      S.L. GLASHOW: ``Partial-symmetry of weak interactions'',
                                Nucl. Phys. 22 (1961) 579;\l
      A. SALAM: ``Weak and electromagnetic interactions'',
             in ``Elementary Particle Theory: Relativistic Groups and
             Analyticity'' (Nobel symposium No 8), edited by N. Svartholm
             (Almquist and Wiksell, Stockholm 1968);\l
      S. WEINBERG: ``A model of leptons'', Phys. Rev. Lett. 19 (1967) 1264.

\bibitem{CabibboKobayashiMaskawa}
       N. CABIBBO: ``Unitary symmetry and leptonic decays'',
                           Phys. Lett. 10 (1963) 513;\l
       M. KOBAYASHI and T. MASKAWA:  ``$CP$-Violation in the Renormalizable
          Theory of Weak Interactions'', Prog. Theor. Phys. 49 (1973) 652.

\bibitem{GellMann}
       M. GELL-MANN: ``A schematic model of baryons and mesons'',
                       Phys. Lett. 8 (1964) 214;\l
       Y. NEEMAN \& M. GELL-MANN: ``The eightfold way'' (Benjamin, N.Y.  1964).

\bibitem{QCD}
       see for example:\l
       W. MARCIANO \& H PAGELS: ``Quantum Chromodynamics'',
            Physics Reports C 36 (1978) 137-276,\l
and references therein.

\bibitem{Machet2}
       B. MACHET: `` An electroweak $SU(2)_L \times U(1)$ gauge theory of
                   $J=0$ mesons'', extended version of the talk ``Custodial
symmetry in a $SU(2)_L \times U(1)$ gauge theory of $J=0$ mesons'' given at
the $2^{nd}$ International Symposium on Symmetries in Subatomic Physics,
Seattle (Washington), June 25th-29th 1997, preprint PAR-LPTHE 97/32,
hep-ph/9709278.

\bibitem{GellMannNishijima}
       M. GELL-MANN: ``Isotopic Spin and New Unstable Particles'',
                                      Phys. Rev. 92 (1953) 833;\l
       K. NISHIJIMA: ``Charge Independence Theory of V particles'',
                                     Prog. Theor. Phys. 13 (1955) 285.

\bibitem{Machet3}
       B. MACHET: ``Some aspects of pion physics in a dynamically broken
                 abelian gauge theory'', Mod. Phys. Lett. A 9 (1994) 3053-3062.

\bibitem{Machet4}
       B. MACHET: ``Comments on the Standard Model of electroweak
                    interactions'', Int. J. Mod. Phys. A 11 (1996) 29-63.

\bibitem{Weinberg}
        S. WEINBERG: ``Current Algebra and Gauge Theories. I'', Phys. Rev.
                              D8 (1973) 605;\l
                ``Current Algebra and Gauge Theories. II. Non-Abelian
                                Gluons'', Phys. Rev. D8 (1973) 4482.

\bibitem{Sikivie}
        P. SIKIVIE, L. SUSSKIND, M. VOLOSHIN and V. ZAKHAROV: ``Isospin
                  breaking in technicolour models'', Nucl. Phys. B 173
                                                            (1980) 189.

\bibitem{CurrentAlgebra}
        see for example:\l
        S.L. ADLER and R.F. DASHEN: ``Current Algebra and Application
                              to Particle Physics'', (Benjamin, 1968);\l
        R. DASHEN: ``Chiral $SU(3) \times SU(3)$ as a Symmetry of Strong
                     Interactions'', Phys. Rev. 183 (1969) 1245;\l
        B.W. LEE: ``Chiral Dynamics'', (Gordon Breach, 1972),\l
and references therein.

\bibitem{Jarlskog}
       C. JARLSKOG: ``$CP$ violation'' (C. Jarlskog ed., World Scientific,
                           Singapore, 1989),\l
and references therein.

\bibitem{Nir}
       Y. NIR: ``$CP$ Violation'', Lectures given at 20th Annual SLAC Summer
            Institute on Particle Physics: The Third Family and the Physics
             of Flavor (School: Jul 13-24, Topical Conference: Jul 22-24,
             Symposium on Tau Physics: Jul 24), Stanford, CA, 13-24 Jul 1992.
             Published in SLAC Summer Inst.1992:81-136 (QCD161:S76:1992);\l
       Y. NIR: ``Recent Developments in Theory of $CP$ Violation'',
             Invited plenary talk given at the 18th International
             Symposium on Lepton Photon Interactions, Hamburg, Germany,
             July 28 - August 1 1997, hep-ph/9709301,\l
and references therein.

\bibitem{NA48-KTEV}
        see for example:\l
        C. TALAMONTI: ``Status report of the NA48 experiment at the
            CERN SPS'', proc. of the 6th Conf. on the Intersection
                      of Particle and Nuclear Physics,
                      Big Sky, Montana (CIPANP97), hep-ex/9708003;\l
        R. BEN-DAVID: ``Status of the KTeV Experiment at Fermilab'',
                 proceedings of the XVI International Workshop on Weak
                 Interactions and Neutrinos, WIN '97,
                 Capri, Italy, June 22-28, 199, hep-ex/9801005.
 
\bibitem{ChristensonCroninFitchTurlay}
       J.H. CHRISTENSON, J.W. CRONIN, J.W. FITCH and R. TURLAY:
           ``Evidence for the $2\pi$ decay of the $K^0_2$ meson'',
             Phys. Rev. Lett. 13 (1964) 138;\l
       V.L. FITCH: ``The discovery of charge-conjugation parity asymmetry'',
             Rev. Mod. Phys. 53 (1981) 367;\l
       J.W. CRONIN: ``$CP$ symmetry violation - the search for its origin'',
                          Rev. Mod. Phys. 53 (1981) 373.

\bibitem{Argus}
       H. ALBRECHT {\em et al.} (ARGUS collaboration): ``Observation of
             $B^0-\ol{B^0}$ mixing'', Phys. Lett. B 245 (1987), 245.

\bibitem{LeeWeinberg}
        T.D. LEE: ``A Theory of Spontaneous $T$ Violation'',
                      Phys. Rev. D 8 (1973) 1226;\l
        S. WEINBERG: ``Gauge Theory of $CP$ Nonconservation'', Phys. Rev.
                      Lett. 37 (1976) 657.

\bibitem{bottom}
        S.W. HERB {\em et al.}: ``Observation of a Dimuon Resonance at
                  $9.5$GeV in $400$GeV Proton Nucleus Collisions'',
                  Phys. Rev. Lett. 39 (1977) 252.

\bibitem{top}
        F. ABE et al.: ``Evidence for top quark production in $\bar p p$
            collision at $\sqrt{s} = 1.8 TeV$'', Phys. Rev. D50 (1994)2966;
        {\em ibidem}: Phys. Rev. Lett. 73 (1994) 225;\l
        S. ABACHI et al.: ``Search for the top quark in $\bar p p$ collisions
                at $\sqrt{s} = 1.8 TeV$'', Phys. Rev. Lett. 72 (1994), 2138.

\bibitem{GUTS}
        see for example:\l
        J. ELLIS: ``Phenomenology of unified gauge theories'', in
                   ``Gauge Theories in High Energy Physics'', lectures at
                    Les Houches Summer School, 1981 (Mary K. Gaillard and
                                         R. Stora eds., North Holland 1983),\l
and references therein.

\bibitem{HORIZONTAL}
        see for example:\l
        H. GEORGI \& S. GLASHOW: ``Attempt to Calculate the Electron Mass'',
              Phys. Rev. D7 (1973) 2457;\l
        S. WEINBERG: ``The problem of mass'', in ``Festschrift f\"ur I.I.
                  Rabi'' ( Lloyd Motz ed., N. Y. Academy of Sciences,
                        New York 1977);\l
        F. WILCZEK \& Z. ZEE: `` Discrete flavour symmetries and a formula
                  for the Cabibbo angle'', Phys. Lett. 70B (1977) 418;\l
        H. FRITZSCH: ``Calculating the Cabibbo angle'', Phys. Lett. 70B
                          (1977) 436;\l
        S.M. BARR \& A. ZEE: `` Calculating the electron mass in terms of
                    measured quantities'', Phys. Rev. D17 (1978) 1854;\l
        F. WILCZEK \& A. ZEE: `` Horizontal Interactions and Weak Mixing
                    Angles'' Phys. Rev. Lett. 42 (1979) 421;\l
        G. ZOUPANOS:``Horizontal interactions as the source of family mixing'',
          Talk given at 1st Hellenic School in Elementary Particle Physics,
          Corfu, Greece, Sep 12-30, 1982 
          (Published in Corfu School 1982:659 (QCD161:H42:1982)),\l
and references therein. 

\bibitem{SUSY}
        see for example:\l
        J. WESS \& J. BAGGER: ``Supersymmetry and Supergravity'' (Princeton
                                        Series in Physics, 1983);\l
        S. POKORSY: ``Weak scale supersymmetry'', lectures at Gif Summer
                                          School (1996),\l
and references therein.

\bibitem{AdlerBellJackiw}
        S.L. ADLER: ``Axial-Vector Vertex in Spinor Electrodynamics'',
                                        Phys. Rev. 177 (1969) 2426;\l
        J.S. BELL and R. JACKIW: ``A PCAC Puzzle: $\pi^0 \rar \gamma\gamma$
                       in the $\sigma$-Model'', Nuovo Cimento 60 (1969) 47;\l
        W.A. BARDEEN: ``Anomalous Ward Identities in Spinor Field
                                  Theories'', Phys. Rev. 184 (1969) 1848.
\bibitem{BouchiatIliopoulosMeyer}
        C. BOUCHIAT, J. ILIOPOULOS and Ph. MEYER: ``An anomaly-free version
              of Weinberg's model'', Phys. Lett. 38 B (1972) 519;\l
        D.J. GROSS and R. JACKIW: ``Effect of Anomalies on
                Quasi-Renormalizable Theories'', Phys. Rev. D 6 (1972) 477.

\bibitem{Quigg}
        see for example:\l
        C. QUIGG: ``Gauge Theories of the Strong, Weak and Electromagnetic
                Interactions'', Frontiers in Physics (Addison Wesley 1994),\l
and references therein.

\bibitem{CHIRALTHEORIES}
        see for example:\l
        J.A. CRONIN: ``Phenomenological Model of Strong and Weak
                       Interactions in Chiral $U(3) \times U(3)$'',
                                            Phys. Rev. 161 (1967) 1483;\l
        J. WESS \& B. ZUMINO: ``Lagrangian Method for Chiral Symmetry'',
                                             Phys. Rev. 163 (1967) 1727;\l
        S. WEINBERG: ``Nonlinear Realizations of Chiral Symmetry'', Phys.
                                                   Rev. 166 (1968) 1568;\l
        S. COLEMAN, J. WESS \& B. ZUMINO: ``Structure of Phenomenological
                               Lagrangian. I'', Phys. Rev. 177 (1969) 2239;\l
        C. CALLAN, S. COLEMAN, J. WESS \& B. ZUMINO: ``Structure of
              Phenomenological Lagrangian. II'', Phys. Rev. 177 (1969) 2247;\l
        S. GASIOROWICZ \& D.A. GEFFEN: ``Effective Lagrangian and Field
            Algebras with Chiral Symmetry'', Rev. Mod. Phys. 41 (1969) 531;\l
        J. GASSER \& H. LEUTWYLER: ``Chiral Perturbation Theory to One
                                    Loop'', Ann. Phys. 158 (1984) 142;\l
        H. LEUTWYLER: ``Chiral effective Lagrangians'', lectures given at
                Theor. Adv. Study Inst., Boulder, 1991, in ``Perspectives in
                 the Standard Model'' (R.K. Ellis, C.T. Hill and J.D.
                Lykken eds., World Scientific, Singapore, 1992),\l
and references therein.

\bibitem{LeeQuiggThacker}
        B.W. LEE, C. QUIGG \& H.B. THACKER: `` Weak interactions at very
                         high energies: The role of the Higgs-boson mass'',
                         Phys. Rev. D 16 (1977) 1519.

\bibitem{sigmamodel}
        M. GELL-MANN \& M. LEVY: ``The Axial Vector Current in Beta Decay'',
                            Nuov. Cim. 16 (1960) 705;\l
        A. D'ADDA, P. di VECCHIA \& M. LUSHER: `` A $1/n$ expandable series
                   of nonlinear $\sigma$ models with instantons'',
                   Nucl. Phys. B146 (1978) 63;\l
                    ``Confinement and chiral symmetry breaking in $CP^{n-1}$
                      models with quarks'', Nucl. Phys. B152 (1979) 125;\l
        A.P. BALACHANDRAN, A. STERN \& G. TRAHERN: ``Nonlinear models as
                    gauge theories'', Phys. Rev. D19 (1979) 2416.

\bibitem{Gatto}
        R. CASALBUONI, D. DOMINICI \& R. GATTO: ``Effective Lagrangian
                   description of the possible strong sector of the
                   Standard Model'' Phys. Lett. B147 (1984) 419.

\bibitem{NambuJonaLasinio}
        Y. NAMBU and G. JONA-LASINIO: ``Dynamical Model
               of Elementary Particles Based on an Analogy with
               Superconductivity'' Phys. Rev. 122 (1961) 345; Phys. Rev. 124
               (1961) 246.

\bibitem{TECHNICOLOUR}
        see for example:\l
        L. SUSSKIND: ``Dynamics of spontaneous symmetry breaking in the
                       Weinberg-Salam theory'', Phys.Rev. D20 (1979) 2619;\l
        S. WEINBERG: ``Implications of dynamical symmetry breaking'', Phys.
                           Rev. D13 (1975) 974, Phys. Rev. D19 (1979) 1277;\l
        G.L. KANE: ``Generalized Higgs physics and technicolour'', in
                   ``Gauge Theories in High Energy Physics'', lectures at
                    Les Houches Summer School, 1981 (Mary K. Gaillard and
                                         R. Stora eds., North Holland 1983),\l
and references therein.

\bibitem{Machet5}
        B. MACHET: ``The Higgs boson and $K \rar \pi\pi$ decays in a
                 $SU(2)_L \times U(1)$ gauge theory of $J=0$ mesons'',
                                  preprint PAR-LPTHE (1998) to appear.

\bibitem{BellonMachet}
        M. BELLON \& B. MACHET: ``The Standard Model of leptons as a purely
                    vectorial theory'', hep-ph/9305212,
                                       Phys. Lett. B 313 (1993) 141.

\bibitem{Regge}
        T. REGGE: ``Introduction to Complex Orbital Momenta'', Nuov. Cim.
                                XIV (1959) 951.

\bibitem{Olive}
         See for example:\l
        D.I. OLIVE: ``Exact electromagnetic duality'', invited talk at the
                      Trieste Conference on Recent Developments in Statistical
                      Mechanics and Quantum Field Theory (April 1995),
                      preprint SWAT/94-95/81 (1995), and references therein.
\bibitem{Cho}
        Y.M. CHO \& D. MAISON: `` Monopole configuration in Weinberg-Salam
                       model'', Phys. Lett. B 391 (1997) 360-365;\l
        Y.M. CHO \& K. KIMM: ``Electroweak Monopoles'', hep-th/9705213;\l
        Y.M. CHO \& K. KIMM: ``Finite Energy Electroweak Monopoles'',
                               hep-th/9707038.

\bibitem{Skyrme}
        T.H.R. SKYRME: ``A nonlinear field theory'',
                                   Proc. Roy. Soc. A260 (1961) 127;\l
        E. WITTEN: ``Global aspects of Current Algebra'', Nucl. Phys. B 223
                 (1983) 422;\l ``Current Algebra, baryons, and quark
                  confinement'', Nucl. Phys. B 223 (1983) 433;\l
        G.S. ATKINS, C.R. NAPPI and E. WITTEN: ``Static properties of the
               nucleon in the Skyrme model'',  Nucl. Phys. B 228 (1983) 552.

\bibitem{GildenerWeinberg}
        E. GILDENER and S. WEINBERG: ``Symmetry breaking and scalar bosons'',
                                     Phys. Rev. D 13 (1976) 3333;\l
        E. GILDENER: ``Gauge-symmetry hierarchies'', Phys. Rev. D 14 (1976)
                                      1667.
%
\end{thebibliography}
\end{document}